\documentclass[aps,prl,twocolumn,floatfix,superscriptaddress,showpacs,showkeys]{revtex4-1}
\usepackage{epsf,amsmath,amssymb,verbatim,color,multirow,pifont,graphicx,tabularx}
\usepackage{braket}
\usepackage{commath}
\usepackage{hyperref}
\usepackage{cleveref}
\usepackage{xparse}
\usepackage{xcolor}

\crefname{equation}{Eq.}{Eqs.}
\crefname{figure}{Fig.}{Figs.}
\crefname{section}{Sec.}{Secs.}
\crefname{chapter}{Chap.}{Chaps.}

\begin{document}
\setcounter{page}{1}
\title[]{Origin of the mixed-order transition in multiplex networks: the Ashkin-Teller model}
\author{S.~\surname{Jang}}
\affiliation{Department of Physics and Chemistry, Korea Military Academy, Seoul 139-804, Korea}
\author{J.S.~\surname{Lee}}
\email{jslee@kias.re.kr}
\affiliation{School of Physics, Korea Institute for Advanced Study, Seoul 130-722, Korea}
\author{S.~\surname{Hwang}}
\affiliation{Department of Physics and Astronomy, Seoul National University, Seoul 151-747, Korea}
\affiliation{Institute for Theoretical Physics, University of Cologne, 50937 K\"oln, Germany}
\author{B.~\surname{Kahng}}
\email{bkahng@snu.ac.kr}
\affiliation{Department of Physics and Astronomy, Seoul National University, Seoul 151-747, Korea}
\date[]{Received \today}

 
\begin{abstract}
Recently, diverse phase transition (PT) types have been obtained in multiplex networks, such as discontinuous, continuous, and mixed-order PTs. However, they emerge from individual systems, and there is no theoretical understanding of such PTs in a single framework. Here, we study a spin model called the Ashkin-Teller (AT) model in a mono-layer scale-free network; this can be regarded as a model of two species of Ising spin placed on each layer of a double-layer network. The four-spin interaction in the AT model represents the inter-layer 
interaction in the multiplex network. 
Diverse PTs emerge depending on the inter-layer coupling strength and network structure. Especially, we find that mixed-order PTs occur at the critical end points. The origin of such behavior is explained in the framework of Landau-Ginzburg theory.
\end{abstract}

\maketitle
Recently, multiplex networks have become a platform for research in network science because in real-world systems, networks are intertangled and function together. Examples include infrastructure in everyday life such as power stations, transportation systems, information networks, and water supply systems.
Multiplex networks can be fragile because failure in one network can cause failure in another, leading to cascading back-and-forth failures. Then, the entire system can exhibit a discontinuous percolation transition ~\cite{Buldyrev2010}.

Mixed-order phase transitions (PTs) (or hybrid PTs) are also observed in multiplex networks \cite{baxter2010}. The size of a giant viable cluster in a multiplex network shows such a PT as a function of the fraction of undamaged nodes. Here, a mixed-order PT means that while the order parameter exhibits a discontinuous PT, the mean cluster size (susceptibility) diverges. Consequently, features of both continuous and discontinuous PTs appear at the same transition point. Mixed-order PTs have been found in several physical models, such as the bootstrap model \cite{leath}, jamming percolation \cite{adler,chayes}, the Ising model with long-range interactions \cite{mukamel}, and the synchronization model \cite{mendes}. However, the mechanism underpinning such mixed-order PTs is not fully understood.

Effort has been made to understand the diverse patterns that emerge from cooperative phenomena in complex networks using spin models in thermal equilibrium. For example, opinion formation and the spread of epidemics have been studied using the Ising ~\cite{ising} and Potts \cite{dslee} models, respectively. Such studies of spin models give some insight into what might happen in complex systems. Motivated by this idea, the Ising model was studied on a double-layer network \cite{holyst}. Interestingly, it exhibited a discontinuous PT, whereas it shows a continuous PT in a mono-layer network. These results led us to speculate that the type of PT can be changed systematically by controlling the inter-layer coupling strength; in the above examples, zero coupling strength (no connection) results in a continuous PT, whereas finite coupling strength leads to a change in PT type.

To investigate the origin of such diverse types of PT in a single theoretical framework systematically, we studied a spin model called the Ashkin–Teller (AT) model \cite{ashkin} in mono-layer scale-free (SF) networks. The AT model contains two species of Ising spin: the $s$-spin and $\sigma$-spin. We want to regard these as a single species of Ising spin placed on the respective layers of a double-layer multiplex network, as shown in Fig.~\ref{fig:double_layer_network}. By controlling the inter-layer interaction, the change in PT type is investigated systematically. Applying Landau-Ginzburg theory to the AT model on SF networks, we obtain a rich phase diagram containing the paramagnetic, Baxter, and so-called $\langle \sigma s \rangle$ phases (in which the product of the $\sigma$ and $s$ spins is ordered, but $\langle s\rangle=\langle \sigma \rangle=0$). The PTs between those phases also include diverse types: continuous, discontinuous, and mixed-order PTs. They meet at tricritical or critical end (CE) points. We find that the mixed-order PT occurs at the CE point.

The AT model contains the two order parameters: the magnetization $m\equiv \langle s \rangle = \langle \sigma \rangle$ and the magnetization of coupled spins $M\equiv \langle \sigma s \rangle$. These are referred to as $m$-magnetization and $M$-magnetization, respectively. These two magnetizations are related, and they generate two singular terms with alternative signs in the Landau-Ginzburg free energy. Competition between these two terms produces diverse profiles in the free energy function. The mixed-order PT emerges when the free energy becomes zero, but the magnetization is finite at the second-order transition temperature.

Let us start by introducing the AT model specifically. Two species of Ising spin, $s_i$ and $\sigma_i$ with states $s_i=\pm 1$ and $\sigma_i = \pm 1$, respectively, are placed at each node $i$, as shown in Fig.~\ref{fig:double_layer_network}. 
The Hamiltonian of the AT model, denoted as $\mathcal{H}_o$, is represented as 
\begin{equation}
-\beta\mathcal{H}_o=K_2\sum_{\langle i,j \rangle}s_is_j
+K_{2}\sum_{\langle i,j\rangle}\sigma_i\sigma_j
+K_4\sum_{\langle i,j\rangle}s_i\sigma_is_j\sigma_j
\label{eq:hamiltonian_beta}
\end{equation}
where $\beta=1/k_{\rm B}T$ with the Boltzmann constant $k_{\rm B}$ and temperature $T$, $K_2=\beta J_2$ and 
$K_4=\beta J_4$ with coupling constants $J_2$ and $J_4$, and $\langle i,j \rangle$ runs over all pairs of nodes connected by links. The SF network is a random graph in which each node has a heterogeneous number of connections, referred to as degree $k$, following the power law $P_d(k)=N_{\lambda} k^{-\lambda}$. 
The coupling between the two layers is shaped in the form of the four-spin interaction with the coupling constant $J_4$.

\begin{figure}
\resizebox{0.99\columnwidth}{!}{\includegraphics{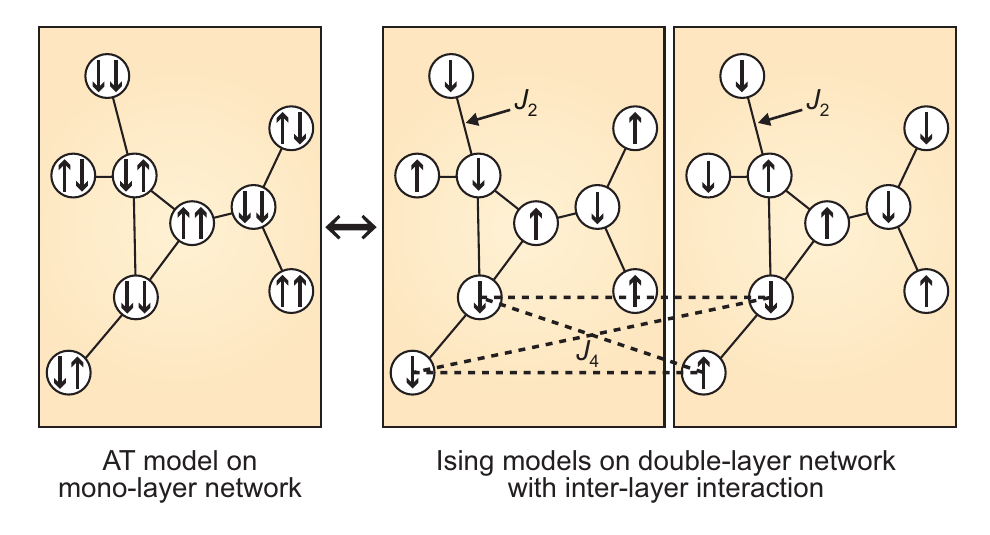}}
\caption{(Color online) The AT model on a mono-layer network may be regarded as a two-species-of-Ising-spin model with an inter-layer interaction (dashed lines) on a double-layer network.} 
\label{fig:double_layer_network}
\end{figure}

\begin{figure}
\resizebox{0.9\columnwidth}{!}{\includegraphics{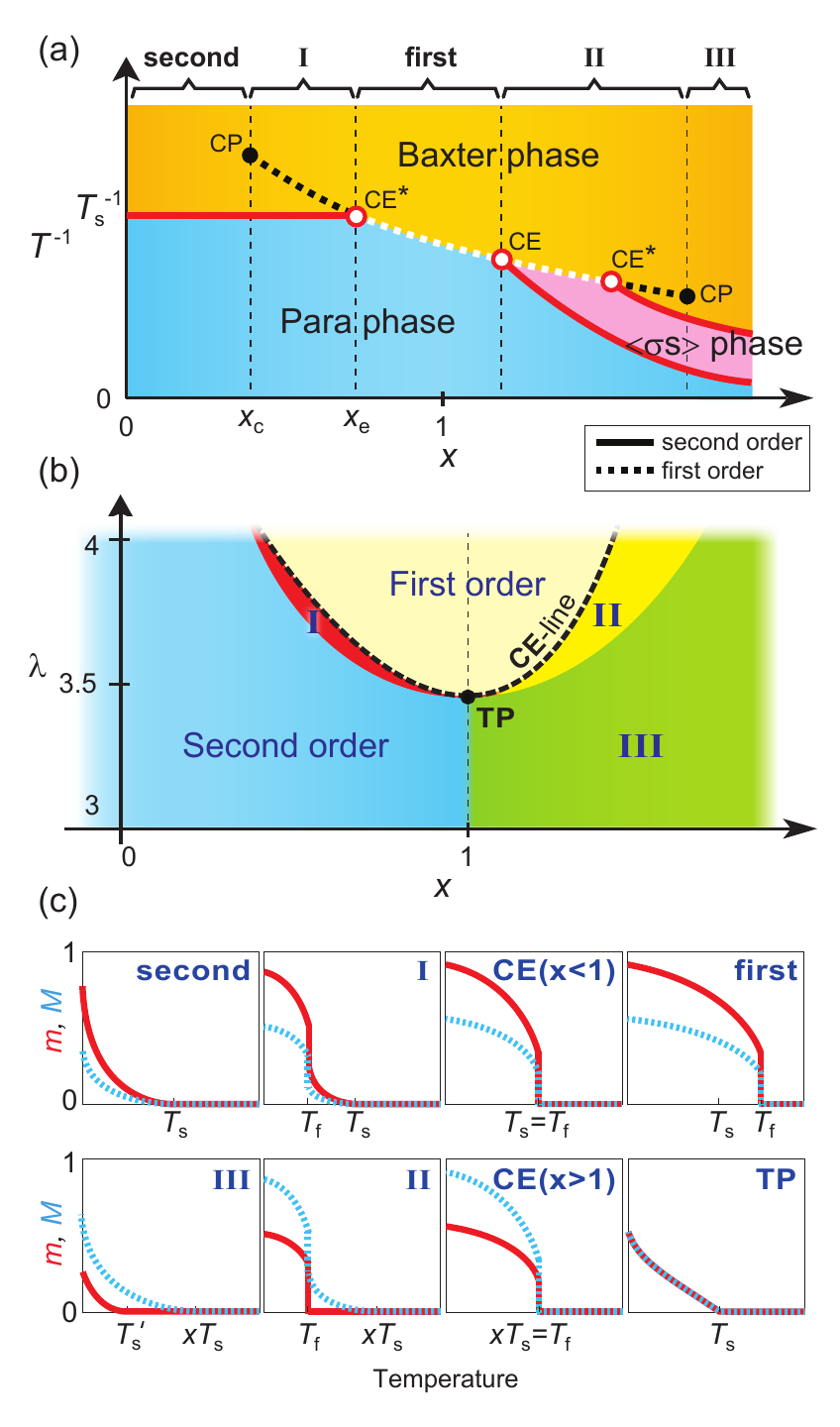}}
\caption{(Color online) Schematic phase diagram in the spaces (a) [$x,T^{-1}$] for $\lambda_c < \lambda < 4$ and (b) [$x,\lambda$] for $3 < \lambda < 4$. 
CP, critical point; CE, critical end point; TP, tricritical point.
The behaviors of the two order parameters $m$ and $M$ in the regimes, {\rm second, I, first, II} and {\rm III} and at the points {\rm CE} and {\rm TP} are shown in (c).} 
\label{fig:phase_diagram}
\end{figure}

The phase diagram of the AT model on a regular lattice in the mean-field limit has been studied extensively \cite{mf}. 
It looks similar to Fig.\ref{fig:phase_diagram}(a) that we obtain for the multiplex SF network, but the first-order lines between the critical point (CP) and CE point are absent. Therefore, the CE points with asterisks in Fig.\ref{fig:phase_diagram}(a) are reduced to a tricritical point in the mean-field solution on a regular lattice \cite{mf}. In SF networks with $3 < \lambda < 4$, owing to those first-order lines, diverse PTs emerge.

Using a standard method, we obtain the mean-field free energy density defined as 
$f \equiv \beta \mathcal{F} / N$, where $\mathcal{F}$ is the free energy defined as 
\begin{align}
f&\simeq -2\int_{k_{\rm min}}^{\infty}\ln\left[
\cosh\left(K_2 m k\right) \right] P_d(k) dk \cr
&~ -\int_{k_{\rm min}}^{\infty}\ln\left[
\cosh\left(K_4 M k\right) \right] P_d(k) dk \cr
&~ -\mathcal{B}_1+K_2 m^2 \langle k\rangle + \dfrac{1}{2}K_4 M^2 \langle k\rangle,
\label{eq:freeEnergyDensity}
\end{align}
where 
\begin{equation}
\mathcal{B}_1 = 
\int_{k_{\rm min}}^{\infty}\ln\left[1+
\tanh^2\left(K_2 m k\right)\tanh\left(K_4 M k\right) \right] P_d(k) dk.
\label{eq:B}
\end{equation}
To obtain the above formula, we used the annealed network approximation, 
$\sum_{\langle i,j \rangle} \mathcal{A}_{ij} \rightarrow \sum_{i\neq j} \frac{k_ik_j}{2N\langle k \rangle} \mathcal{A}_{ij}$, 
where $N$ is the total number of nodes, $\langle k \rangle$ is the mean degree of a network, and $\mathcal{A}_{ij}$ is a given function of $i$ and $j$. 
We also used the global magnetization $m_\alpha$ as $m_\alpha= {\sum_{i}k_i m_\alpha^i}/{N\langle k\rangle}$, where $m_\alpha^i$ is local order parameter. Then, we set $m_s = m_\sigma \equiv m$ and $m_{s\sigma}\equiv M$.

The minimization conditions $\frac{\partial f}{\partial m}=0$ and $\frac{\partial f}{\partial M}=0$ lead to the following self-consistent relations:
\begin{equation}
m\langle k \rangle =
\int_{k_{\rm min}}^{\infty}
\dfrac{\tanh\left(K_2 m k\right)\left[1+\tanh\left(K_4 m k\right)\right]}
{1+\tanh^2\left(K_2 m k\right)\tanh\left(K_4 M k\right)}
k P_d (k) dk
\label{eq:equationOfState_m}
\end{equation}
and
\begin{equation}
M\langle k \rangle =
\int_{k_{\rm \rm min}}^{\infty}
\dfrac{\tanh\left(K_4 M k\right)+\tanh^2\left(K_2 m k\right)}
{1+\tanh^2\left(K_2 m k\right)\tanh\left(K_4 M k\right)}
k P_d(k) dk.
\label{eq:equationOfState_M}
\end{equation}
There exist three possible solutions for Eqs.~(\ref{eq:equationOfState_m}) and (\ref{eq:equationOfState_M}): the para ($m=M=0$), Baxter ($m, M>0$), and $\langle \sigma s\rangle$ ($m=0, M>0$) phases. $m>0, M=0$ cannot satisfy the above relations. 

To obtain the susceptibility, we consider an AT Hamiltonian with an external field, which is written as 
\begin{equation}
-\beta\mathcal{H}=-\beta\mathcal{H}_o+\sum_{i} k_iH_2 (s_i+\sigma_i)+
\sum_{i} k_iH_4 s_i\sigma_i,
\label{eq:AT_hamiltonian_ext_field}
\end{equation}
where the external fields $H_2$ and $H_4$ are weighted by the degree of each node. Then, the relations between the free energy and magnetization hold $- \partial f /\partial H_2 = m \langle k \rangle$ and $- \partial f /\partial H_4 = M \langle k \rangle$. 
Next, by taking the partial derivative of $m$ with respect to $H_2$ and then taking the limit $H_2, H_4 \rightarrow 0$, we obtain the susceptibilities, $\chi_m \equiv {\partial m}/{\partial H_2}|_{H_2,H_4 \to 0}$ and $\chi_M \equiv {\partial M}/{\partial H_4}|_{H_2,H_4 \to 0}$.

When $K_4=0$, the AT model reduces to the Ising model. In this case, the system is ordered for all temperatures for $2< \lambda \le 3$.
The AT model behaves similarly for this range of degree exponent. Thus, we consider only the case $\lambda >3$. When $x\equiv K_4/K_2=1$, the AT model shows a particular feature. Consequently, we consider this case first and then the case $x\ne 1$. 

i) When $x=K_4/K_2=1$, the two species of spin are indistinguishable. Then, $m \sim M$ and the AT model is reduced to the four-state Potts model \cite{Potts}. Expanding the free energy density~(\ref{eq:freeEnergyDensity}) up to the third order in $m$ gives
\begin{eqnarray}
f&\simeq& \frac{3}{2} K_2 m^2 \langle k\rangle \left(1-\frac{K_2 \langle k^2\rangle}{\langle k\rangle} \right)+ 
\left( 3 C_1 -C_2\right)(K_2 m)^{\lambda-1} \cr &+& \frac{N_\lambda k_\textrm{min}^{4-\lambda}}{4-\lambda}(K_2 m)^3
+\textrm{higher orders (h.o.)}.
\label{eq:freeEnergyDensity_x1}
\end{eqnarray}
Note that both $C_1(\lambda)$ and $C_2(\lambda)$ are positive.
When $3C_1-C_2>0$, which occurs for $\lambda < \lambda_c\approx 3.503$, the second-order PT occurs at $T_s\equiv J_2\langle k^2 \rangle/k_B \langle k \rangle$ with the critical exponent $\beta_m=1/(\lambda-3)$. When $3C_1-C_2 <0$ for $\lambda > \lambda_c$, the first-order PT occurs at $T_f$ $(>T_s)$. $T_f$ is the point at which the free energy becomes globally minimum at a finite $m$ discontinuously.
When $3C_1-C_2=0$ at $\lambda_c$, the continuous transition occurs at $T_s$, but the critical exponent differs as $\beta_\textrm{TP} =1$.
The susceptibility at the TP is obtained, which diverges at both $T_s^+$ and $T_s^-$ with the critical exponent $\gamma_\textrm{TP}=1$. This is the conventional tricritical point (TP), as shown in the panel labeled {\rm TP} in Fig.~\ref{fig:phase_diagram}(c).

ii) When $x=K_4/K_2 \neq 1$, Eqs.(\ref{eq:equationOfState_m}) and (\ref{eq:equationOfState_M}) can be expanded in terms of $m$ and $M$ as follows: 
\begin{eqnarray}
m \langle k\rangle \left(1- \frac{ T_s}{T} \right) &\simeq& C_3 (K_2 m)^{\lambda-2}
+ \mathcal{B}_2+{\rm h.o.}, \label{eq:small_m} \\
M \langle k\rangle \left(1- \frac{ xT_s}{T} \right) &\simeq& C_4 (K_4 M)^{\lambda-2}
+ C_5 (K_2 m)^{\lambda-2}+{\rm h.o.},
\label{eq:small_M}
\end{eqnarray}
where $C_3(\lambda,r_0) <0$, $C_4(\lambda,r_0) <0$, and $C_5(\lambda,r_0) >0$ are numbers of $\mathcal{O}(1)$ and 
they depend on $\lambda$ and $r_0\equiv K_4M/K_2m$. 
$\mathcal{B}_2(K_2 m, K_4 M, \lambda) >0$ and its order depends on the ratio $M/m$. 
These expansions are possible near a continuous transition point, in which $0< m, M \ll 1$. 
Fig.\ref{fig:phase_diagram}(a) and (b) show schematic phase diagrams in the parameter space $(x, T^{-1})$ for $\lambda_c < \lambda < 4$ and $(x,\lambda)$ for $3 < \lambda < 4$, respectively, obtained based on the criteria discussed below and numerical data. 

ii-1) For $x < 1$: As temperature is decreased from a sufficiently large value, the second-order transition for the $m$-magnetization takes place first at $T_s$ because $T_s > xT_s$. 
Near $T_s$, $(1-xT_s/T)$ in Eq.(\ref{eq:small_M}) are a number of $\mathcal{O}(1)$, and so $M$ and $M^{\lambda-2}$ for $\lambda >3$ cannot be of the same order. Instead, $M$ and $m^{\lambda-2}$ should be of the same order, and they are related as 
\begin{align}
M\simeq \frac{C_5^\prime}{\langle k\rangle (1- xT_s/T) } (K_2 m)^{\lambda-2},
\label{eq:M_and_m_relation}
\end{align}
where $C_5^\prime$ is the value of $C_5$ in the limit $M/m \rightarrow 0$, and it depends on $\lambda$. 
In this case, $\mathcal{B}_2$ is of higher order than $m^{\lambda-2}$, and so the magnetization is obtained as $m \sim (T_s -T)^{1/(\lambda-3)}$. Therefore, the critical exponent $\beta_m=1/(\lambda-3)$. 
$M \sim (T_s -T)^{\beta_M}$ with $\beta_M = (\lambda-2)/(\lambda-3)$. 

Next, using Eq.~(\ref{eq:M_and_m_relation}), we expand the free energy~(\ref{eq:freeEnergyDensity}) up to the three lowest-order terms with respect to $m$ as
\begin{align}
f(m)&\simeq K_2 m^2 \langle k\rangle \left(1-\frac{T_s}{T}\right) +2C_1(K_2 m)^{\lambda-1} \cr
&~-\frac{K_4 {C_5^\prime}^2}{2\langle k\rangle (1-\frac{xT_s}{T})}(K_2 m)^{2(\lambda-2)}+\textrm{h.o.}. 
\label{eq:free_energy_x<1}
\end{align}
The $(\lambda-1)$-order term is always positive, whereas the $2(\lambda-2)$-order term is negative. 
Competition between the magnitudes of these two singular terms produces an interesting phase diagram for $3 < \lambda <4$, actually for $\lambda_c < \lambda <4$. 
Note that the coefficient of the $2(\lambda-2)$-order term varies depending on $T$, $x$, and $\lambda$. 

\begin{figure}
\resizebox{0.90\columnwidth}{!}{\includegraphics{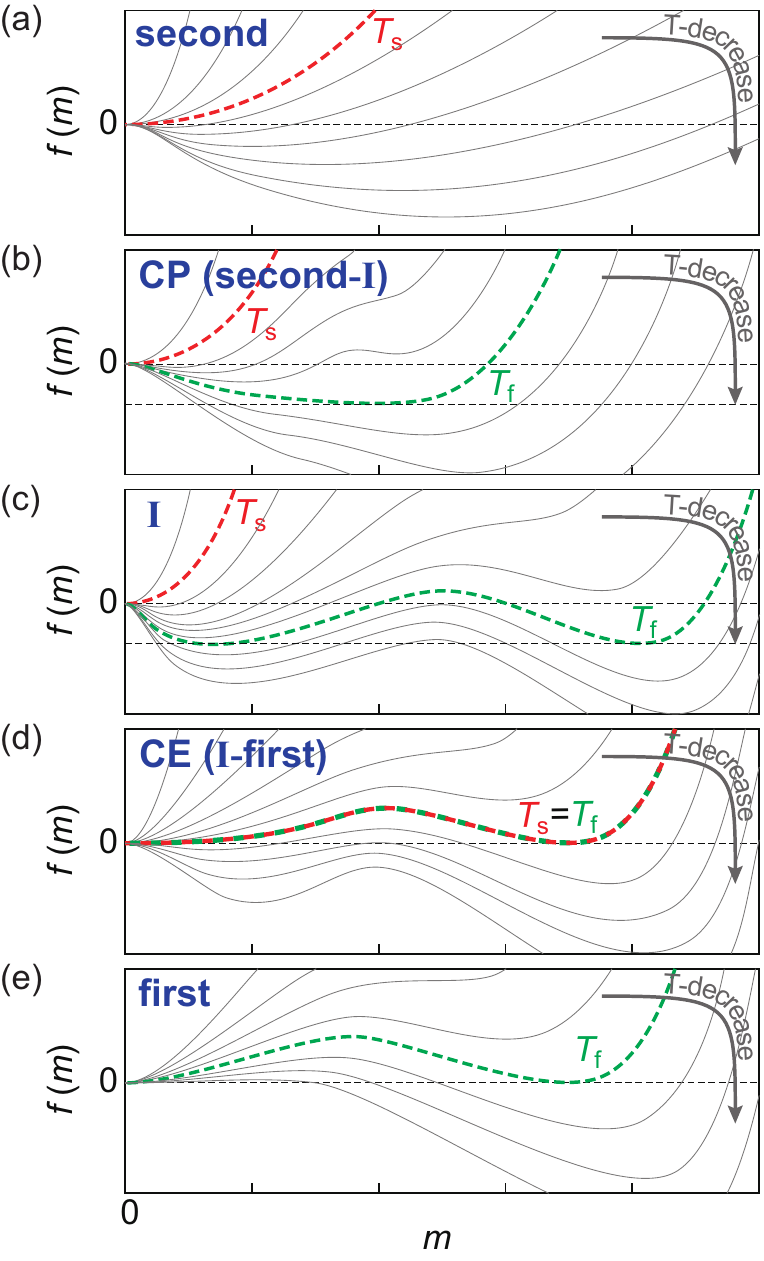}}
\caption{(Color online) (a) Plots of free energy as a function of $m$ for (a) the second-order transition, (b) continuous transition at the critical point, (c) successive transitions in the intermediate regime {\rm I}, (d) discontinuous transition at the CE point, and (e) the first-order transition.
These behaviors appear as temperature varies, but for a given $\lambda$ in $\lambda_c\approx 3.5 <\lambda<4$.} 
\label{fig:free_energy_landscape}
\end{figure}

When $\lambda_c < \lambda <4$, the phase diagram is insensitive to $\lambda$. 
Thus, we consider the free energy as a function of $x$ and $T$. 
First, we sketch the behavior of $f(m)$ vs. $m$ for different values of $x$. 
When $x$ is close to zero, the $2(\lambda-2)$-order term is negligible compared with the $(\lambda-1)$-order term, and the global minimum is located at $m=0$ for $T\le T_s$ (Fig.\ref{fig:free_energy_landscape}a). 
When $T$ is decreased below $T_s$, the $m$-position of the global minimum increases continuously, which leads to a continuous transition. 
On the other hand, when $x$ is close to one, the $2(\lambda-2)$-order term can make a global minimum of $f(m)$ at a certain finite $m\equiv m_2$ for $T \le T_f$, where $T_f >T_s$.
Then, a discontinuous transition takes place at $T_f$ (Fig.\ref{fig:free_energy_landscape}e). 

In the intermediate regime {\rm I}$=[x_c,x_e]$, as temperature is decreased across $T_s$, the free energy exhibits a global minimum at $m_1(T)> 0$, which increases continuously as the temperature is lowered further. 
In the meantime, a local minimum of $f(m)$ develops at a certain $m_2(T) > m_1(T)$ due to the $2(\lambda-2)$-order term. As the temperature is lowered further beyond a certain temperature, denoted as $T_f$ ($<T_s$), the local minimum at $m_2$ becomes a global minimum, as depicted in Fig.~\ref{fig:free_energy_landscape}(c). That is, $f(m_2(T_f^-)) < f(m_1(T_f^-)) <0$. Accordingly, the magnetization jumps from $m_1$ to $m_2$ discontinuously at $T_f$. Thus, the system exhibits a discontinuous transition at $T_f$. Such successive PTs occur in the intermediate regime of $x$. 

Next, we consider particular points at the boundaries of the regime {\rm I} in Fig.\ref{fig:phase_diagram}(a). The first-order line terminates at a certain point $x_c$, which is called the critical point, as seen in a liquid-gas system. At this point, the transition is continuous, and the behavior of $f(m)$ is shown in Fig.~\ref{fig:free_energy_landscape}(b). On the other hand, the second-order line terminates at a point called the CE point, where the first-order line continues into the regime $x> x_e$. 
This CE point occurs at temperature $T_s$, at which there exist two minima in the free energy function at $m=0$ and $m_2$ ($m_2 >0$), but $f(m)=f(m_2)=0$ as shown in Fig.~\ref{fig:free_energy_landscape}(d). Thus, $T_f=T_s$. 
At $T_s^+$, the magnetization is zero, but at $T_s^-$, which is equivalent to $T_f^-$, the magnetization jumps to $m_2 >0$, and the system shows a discontinuous PT. 

We obtain the susceptibility near $T_s$ as 
 \begin{equation}
\chi_m = \left\{ \begin{array}{cc}
 \frac{T\langle k^2 \rangle}{\langle k \rangle} (T-T_s)^{-1} &~~\textrm{for}~ T>T_s, \\
 \frac{T\langle k^2 \rangle}{\langle k \rangle(\lambda-3)} (T_s-T)^{-1} &~~\textrm{for}~ T<T_s. 
\end{array} \right. 
\label{eq:susceptibility_m_x<1}
\end{equation}
Thus, the susceptibility exponent is obtained as $\gamma_m=1$. This result is valid near any point along the second-order transition line, but not at the CE point. At CE, $\chi_m$ diverges as Eq.~(\ref{eq:susceptibility_m_x<1}) for $T_s^+$ and becomes finite for $T_s^-$ (see \cite{sm}).
Because the $m$-magnetization is discontinuous at the CE point, a mixed-order transition emerges at the CE point.

The susceptibility of the $M$-magnetization is obtained to be finite as 
\begin{equation}
\chi_M = \frac{T \langle k^2 \rangle}{\langle k \rangle (T-xT_s)}, 
\label{eq:susceptibility_M_x<1}
\end{equation}
near $T_s$, although the $M$-magnetization exhibits a continuous PT at $T_s$. Thus, the $M$-magnetization exhibits another type of mixed-order PT at the continuous transition point for $0<x<1$.

Consider the phase diagram in the space $(x,\lambda)$ for $3 < \lambda < 4$ (Fig.\ref{fig:phase_diagram}b). 
As $\lambda \to \lambda_c^+$, the regimes $\rm I$ and $\rm II$ of the successive transitions shrink to the tricritical point at $x=1$ and $\lambda_c$. Remarkably, the upper boundary of regimes {\rm I} and {\rm II} becomes a critical end line on which mixed-order PTs take place. 


ii-2) When $x>1$, besides the para and Baxter phases, the $\langle \sigma s \rangle$ phase also exists. As shown in Fig.~\ref{fig:phase_diagram}(a), there exist three regimes: in the first-order transition regime near $x=1^+$, a discontinuous PT occurs at $T_f$ from the para to the Baxter phase, as depicted in panel {\rm ``first"} of Fig.~\ref{fig:phase_diagram}(c). The difference is that $M$ is larger than $m$. In the regime {\rm III} for $x \gg 1$, $M$ and $m$ undergo the second-order PT at different temperatures, $xT_s$ and $T_s^\prime(=T_s/(1-C_6))$ (see \cite{sm} for $C_6$), respectively, as shown in panel {\rm III} in Fig.~\ref{fig:phase_diagram}(c). In the intermediate regime, the $\langle \sigma s\rangle$ phase exists between $xT_s$ and $T_s^\prime$, and more than one PT occurs successively.
Note that the critical exponent for both magnetizations $m$ and $M$ is $\beta_m=\beta_M=1/(\lambda-3)$. Mixed-order PTs also occur at the CE points.

In this Letter, we studied the Ashkin–Teller model on a mono-layer scale-free network using the mean-field approximation. We obtained a rich phase diagram containing diverse types of phase transition, such as second-order, first-order, and mixed-order transitions, and diverse types of transition point, such as critical, CE, and tricritical points. Particularly, there exist the CE points as in liquid $^3{\rm He}-^4{\rm He}$ mixtures \cite{superfluid} and metamagnets \cite{metamagnet}, at which the mixed-order transitions emerge in the AT model we studied. The rich phase diagram is created as collective phenomena of spins for the asymmetric case $(x\ne 1)$ between the intra- and inter-layer interaction strengths. Note that the CE points do not exist but are reduced as a tricritical point for the symmetric case $(x=1)$. We anticipate that such a rich phase diagram obtained in thermal equilibrium systems could be a guideline for understanding complex phenomena in multi-layer networked systems.

This research was supported by the NRF grant Nos.~2011-35B-C00014 (JSL) and 2010-0015066 and SNU R\&D grant (BK). \\

\onecolumngrid
\newpage
\appendix
\section{Supplementary Material}

\subsection{Numerical test of the order parameter curves for respective regions}
Figure S1 shows the order parameters $m$ and $M$ obtained numerically from the mean-field free energy (see Eq.(2) in the main text). They show the curves for the first, second, TP, I, II, and III regions.
\begin{figure}[b]
\resizebox{0.8\columnwidth}{!}{\includegraphics{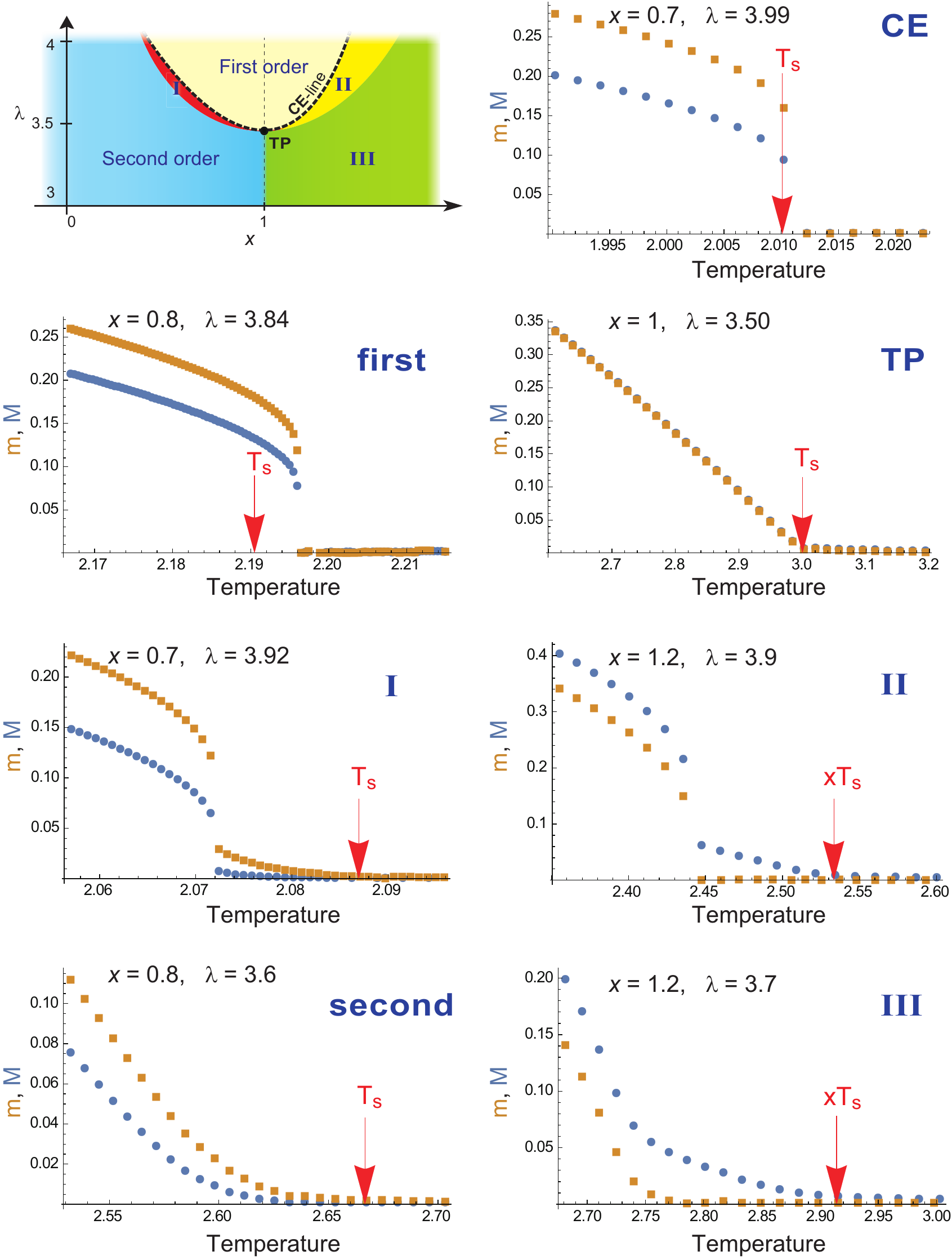}}
\caption{[color online] Plots of the order parameters $m$ and $M$ as a function of temperature in each regime of phase transitions, which are obtained from  different values of $x=K_4/K_2$ and the degree exponent $\lambda$.} 
\label{fig:variousPT}
\end{figure}

\subsection{Definitions for coefficients used in the main text}

\begin{align}
&C_1 (\lambda)= N_\lambda \int_0^\infty \left(-\ln [\cosh y] +\frac{1}{2} y^2 \right) y^{-\lambda},\cr
&C_2 (\lambda) = N_\lambda \int_0^\infty \ln \left[ 1+\tanh^3 y \right]y^{-\lambda} dy,\cr
&C_3 (\lambda,r_0)=N_\lambda \int_0^\infty \left[\frac{\tanh y}{1+\tanh^2 y \tanh(r_0 y)}-y \right] y^{1-\lambda} dy,\cr
&C_4 (\lambda,r_0)=N_\lambda \int_0^\infty \left[\frac{\tanh y}{1+\tanh y \tanh^2(y/r_0)}-y \right] y^{1-\lambda} dy,\cr 
&C_5 (\lambda,r_0)=N_\lambda \int_0^\infty \frac{\tanh^2 y  }{1+\tanh^2 y \tanh(r_0 y)}y^{1-\lambda} dy,\cr 
&C_5^\prime (\lambda) =C_5(\lambda,0)=N_\lambda \int_0^\infty \tanh^2 y  ~ y^{1-\lambda} dy,\cr 
&C_6 = \frac{N_\lambda K_2}{\langle k\rangle} \int_{k_\textrm{min}}^\infty \tanh (K_4 M k) k^{2-\lambda} dk,\cr
&\mathcal{B}_2=N_\lambda \int_{k_\textrm{min}}^\infty \frac{\tanh (K_2 m k) \tanh (K_4 M k)}{1+\tanh^2 (K_2 m k) \tanh(K_4 M k)} k^{1-\lambda} dk.
\end{align}

\subsection{Susceptibility near a continuous transition point}

Consider a AT model Hamiltonian including terms with external field which is a function of degree of a node as
\begin{equation}
-\beta\mathcal{H}=K_2\sum_{\langle i,j \rangle}s_is_j
+K_{2}\sum_{\langle i,j\rangle}\sigma_i\sigma_j
+K_4\sum_{\langle i,j\rangle}s_i\sigma_is_j\sigma_j 
+\sum_{i} h_2(k_i) s_i+ \sum_{i} h_2(k_i) \sigma_i+ \sum_{i} h_4(k_i) s_i\sigma_i.
\label{eq:AT_hamiltonian_ext_field}
\end{equation}
Here, we set $h_2(k_i)$ and $h_4(k_i)$ as the external field weighted by degree of a node as $h_2(k_i) = H_2 k_i$ and $h_4(k_i) = H_4 k_i$, respectively. Then, we have usual free energy and magnetization relation as $- \partial f /\partial H_2 = m \langle k \rangle$ and $- \partial f /\partial H_4 = M \langle k \rangle$. Physically, such weighted external field can be interpreted as follows; one node is driven by same amount of influence through each link, thus, total driving force is proportional to the degree of the node. Following the same derivation in the main text, the free energy for the above Hamiltonian is given by
\begin{align}
f&\simeq -2\int_{k_{\rm min}}^{\infty}\ln\left[
\cosh\left(K_2 m k +H_2 k\right) \right] P_d(k) dk  -\int_{k_{\rm min}}^{\infty}\ln\left[
\cosh\left(K_4 M k +H_4 k\right) \right] P_d(k) dk \cr
&~~~ -\mathcal{B}_1(K_2 m+H_2 k,K_4 M +H_4 k,\lambda)+K_2 m^2 \langle k\rangle + \dfrac{1}{2}K_4 M^2 \langle k\rangle .
\end{align}
The self-consistent relations for $m$ and $M$ are given as follows:
\begin{equation}
m\langle k \rangle =
\int_{k_{\rm min}}^{\infty}
\dfrac{\tanh\left(K_2 m k +H_2 k\right)\left[1+\tanh\left(K_4 M k +H_4 k\right)\right]}
{1+\tanh^2\left(K_2 m k+H_2 k\right)\tanh\left(K_4 M k+H_4 k\right)}
k P_d (k) dk
\label{eq:eqnOfState_m}
\end{equation}
and
\begin{equation}
M\langle k \rangle =
\int_{k_{\rm \rm min}}^{\infty}
\dfrac{\tanh\left(K_4 M k +H_4 k\right)+\tanh^2\left(K_2 m k+H_2 k\right)}
{1+\tanh^2\left(K_2 m k+H_2 k\right)\tanh\left(K_4 M k+H_4 k\right)}
k P_d(k) dk.
\label{eq:eqnOfState_M}
\end{equation}

First, consider the case $x<1$. Equation~(\ref{eq:eqnOfState_m}) can be expanded as
\begin{eqnarray}
m \langle k\rangle \simeq (K_2 m+H_2) \langle k^2 \rangle +C_3\left(\lambda,\frac{K_4 M +H_4}{K_2 m +H_2}\right) (K_2 m+H_2)^{\lambda-2} .
\end{eqnarray}
By taking partial derivative of the above equation in terms of $H_2$ and then taking $H_2, H_4 \rightarrow 0$ limit, we have
\begin{eqnarray}
\chi_m \langle k\rangle \simeq (K_2 \chi_m+1) \langle k^2 \rangle +C_3\left(\lambda,\frac{K_4 M}{K_2 m}\right)(\lambda-2)K_2\chi_m(K_2 m)^{\lambda-3},
\end{eqnarray}
where $\chi_m$ is a susceptibility of $m$ and defined as $\left. \frac{\partial m}{\partial H_2} \right|_{H_2, H_4\rightarrow 0}$.
When the second order phase transition occurs at $T_s$, $m=0$ for $T>T_s$ and $C_3\left(\lambda,\frac{K_4 M}{K_2 m}\right) (K_2m)^{\lambda-3} \approx \frac{\langle k\rangle}{K_2} (1-T_s/T)$ for $T<T_s$ near $T_s$. Then $\chi_m$ becomes
\begin{equation}
\chi_m = \left\{ \begin{array}{cc}
  \frac{T\langle k^2 \rangle}{\langle k \rangle} (T-T_s)^{-1} &~~\textrm{for}~ T>T_s \\
  \frac{T\langle k^2 \rangle}{\langle k \rangle(\lambda-3)} (T_s-T)^{-1} &~~\textrm{for}~ T<T_s 
\end{array} \right. ~~~\textrm{when  }x<1.
\label{eq:susceptibility_m_x<1}
\end{equation}
Similarly, Eq.~(\ref{eq:eqnOfState_M}) can be expanded as 
\begin{eqnarray}
&M \langle k\rangle \simeq (K_4 M+H_4) \langle k^2\rangle + C_4\left(\lambda,\frac{K_2 m +H_2}{K_4 M +H_4}\right) (K_4 M+H_4)^{\lambda-2}
+ C_5\left(\lambda,\frac{K_4 M +H_4}{K_2 m +H_2}\right) (K_2 m +H_2)^{\lambda-2}.~~~~~~~~~
\label{eq:small_M}
\end{eqnarray}
Taking partial derivative of the above equation by $H_4$ and then taking $H_2, H_4 \rightarrow 0$ limit give
\begin{equation}
\chi_M \langle k\rangle \simeq (K_4 \chi_M+1) \langle k^2\rangle + C_4\left(\lambda,\frac{K_2 m}{K_4 M}\right)(\lambda-2)K_4 \chi_M (K_4 M)^{\lambda-3}
+ C_5\left(\lambda,\frac{K_4 M}{K_2 m}\right)K_2 \left. \frac{\partial m}{\partial H_4} \right|_{H_2, H_4\rightarrow 0} (K_2 m )^{\lambda-3},~~~~~~~~~
\end{equation}
where $\chi_M$ is a susceptibility of $M$ and defined as $\left. \frac{\partial M}{\partial H_4} \right|_{H_2, H_4\rightarrow 0}$.
When the second order phase transition occurs at $T_s$, $m=M=0$ for $T>T_s$ and $(K_2m)^{\lambda-3}\sim (1-T_s/T)$ and $M\sim m^{\lambda-2}$ for $T<T_s$ near $T_s$. Then $\chi_M$ becomes
\begin{equation}
\chi_M = \frac{T \langle k^2 \rangle}{\langle k \rangle (T-xT_s)}~~~\textrm{near }T_s.
\label{eq:susceptibility_M_x<1}
\end{equation}
Note that at the critical end point, where transition is discontinuous, the susceptibility diverges for $T_s^+$, whereas it is finite for $T_s^-$. For more information, see the next section.

Now, consider the case $x>1$. When the second order phase transition from para to $\langle \sigma s\rangle$ phase occurs at $T=xT_s$, $m$ is equal to $0$ and Eq.~(\ref{eq:eqnOfState_M}) is expanded as
\begin{eqnarray}
M \langle k\rangle \simeq (K_4 M+H_4) \langle k^2\rangle + C_4\left(\lambda,\frac{H_2}{K_4 M +H_4}\right) (K_4 M+H_4)^{\lambda-2}.
\end{eqnarray}
Taking partial derivative of the above equation by $H_4$ and then taking $H_2, H_4 \rightarrow 0$ limit give
\begin{eqnarray}
\chi_M \langle k\rangle \simeq (K_4 \chi_M+1) \langle k^2\rangle + C_4\left(\lambda,0\right) K_4 \chi_M (\lambda-2)(K_4 M)^{\lambda-3}.
\end{eqnarray}
Using $M=0$ for $T>xT_s$ and $C_4(\lambda,0) (K_4 M)^{\lambda-3}=\frac{\langle k \rangle}{K_4}(1-xT_s/T)$ for $T<xT_s$, $\chi_M$ becomes
\begin{equation}
\chi_M = \left\{ \begin{array}{cc}
  \frac{T\langle k^2 \rangle}{\langle k \rangle} (T-xT_s)^{-1} &~~\textrm{for}~ T>xT_s \\
  \frac{T\langle k^2 \rangle}{\langle k \rangle(\lambda-3)} (xT_s-T)^{-1} &~~\textrm{for}~ T<xT_s 
\end{array} \right. ~~~\textrm{when  }x>1.
\label{eq:susceptibility_M_x>1}
\end{equation}
Therefore, at the critical end point existing at the boundary between `first' and `II' regions, where $xT_s=T_f$, the susceptibility $\chi_M$ diverges for $T-T_s \rightarrow 0^+$ case, though the transition is discontinuous. 
When the second order phase transition from $\langle \sigma s\rangle$ to Baxter phase occurs at $T=T_s/(1-C_6)$, following the same derivation, $\chi_m$ becomes
\begin{equation}
\chi_m = \left\{ \begin{array}{cc}
  \frac{T\langle k^2 \rangle}{\langle k \rangle} \left(T-\frac{T_s}{1-C_6} \right)^{-1} &~~\textrm{for}~ T>T_s/(1-C_6) \\
  \frac{T\langle k^2 \rangle}{\langle k \rangle(\lambda-3)} \left(\frac{T_s}{1-C_6} -T \right)^{-1} &~~\textrm{for}~ T<T_s/(1-C_6) 
\end{array} \right. ~~~\textrm{when  }x>1.
\label{eq:susceptibility_M_x>1}
\end{equation}

\subsection{Susceptibility at the critical end point} 
Since the transition is discontinuous at the critical end point, the expansions with the assumption $m,M \ll 1$ used in the previous section cannot be performed to calculate the susceptibility. Instead, we should keep the explicit integral forms as follows. If we take a derivative of Eq.(\ref{eq:eqnOfState_m}) with respect ot $H_2$ and take $H_2,H_4 \rightarrow 0$ limit, we obtain
\begin{equation}
\chi_m = \frac{\mathcal{A}_1 +\mathcal{A}_2 K_4 \left. \frac{\partial M}{\partial H_2} \right|_{H_2,H_4\rightarrow 0}}{\langle k \rangle \left( 1 - \frac{K_2}{\langle k\rangle} \mathcal{A}_1 \right)}
\label{chi_m_explicit}
\end{equation}
where
\begin{equation}
\mathcal{A}_1 = N_\lambda \int_{k_\textrm{min}}^\infty \frac{[1-\tanh^2 (K_2 m k) \tanh (K_4 M k)][1+\tanh (K_4 M k)]}{[1+\tanh^2 (K_2 m k) \tanh (K_4 M k)]^2 \cosh^2 (K_2 m k)} k^{2-\lambda}dk ,
\end{equation}
and
\begin{equation}
\mathcal{A}_2 = N_\lambda \int_{k_\textrm{min}}^\infty \frac{[1-\tanh^2 (K_2 m k)]\tanh (K_2 m k)}{[1+\tanh^2 (K_2 m k) \tanh (K_4 M k)]^2 \cosh^2 (K_4 M k)} k^{2-\lambda}dk .
\end{equation}
To evaluate Eq.~(\ref{chi_m_explicit}), we also should calculate $\left. \frac{\partial M}{\partial H_2} \right|_{H_2,H_4\rightarrow 0}$. If we take a derivative of Eq.~(\ref{eq:eqnOfState_M}) with respect to $H_2$ and take $H_2,H_4 \rightarrow 0$ limit, we obtain
\begin{equation}
\left. \frac{\partial M}{\partial H_2} \right|_{H_2,H_4\rightarrow 0} = \frac{\mathcal{A}_4 \left( \chi_m K_2 +1 \right)}{\langle k \rangle \left( 1 - \frac{ K_4}{\langle k\rangle} \mathcal{A}_3 \right)}
\label{chi_mm_explicit}
\end{equation}
where
\begin{equation}
\mathcal{A}_3 = N_\lambda \int_{k_\textrm{min}}^\infty \frac{1-\tanh^4 (K_2 m k) }{[1+\tanh^2 (K_2 m k) \tanh (K_4 M k)]^2 \cosh^2 (K_4 M k)} k^{2-\lambda}dk ,
\end{equation}
and
\begin{equation}
\mathcal{A}_4 = N_\lambda \int_{k_\textrm{min}}^\infty \frac{2\tanh(K_2 m k)[1-\tanh^2 (K_4 M k)] }{[1+\tanh^2 (K_2 m k) \tanh (K_4 M k)]^2 \cosh^2 (K_2 m k)} k^{2-\lambda}dk .
\end{equation}
At the critical end point, $m=M=0$ for $T>T_s$, so $\mathcal{A}_1 =\langle k^2\rangle$ and $\mathcal{A}_2=0$. Then, $\chi_m$ becomes the same as the result in Eq.~(\ref{eq:susceptibility_m_x<1}) for $T>T_s$ case. For $T<T_s$, $\chi_m$ can be numerically evaluated by solving Eqs.~(\ref{chi_m_explicit}) and (\ref{chi_mm_explicit}) together. Figure S2 shows the numerically calculated $\chi_m$ near $T_s$ at the critical end point with $\lambda=3.99$ and $x=0.7$, which is the case of CE  in Fig. S1. It clearly shows that the susceptibility diverges for $T>T_s$, whereas it becomes finite for $T<T_s$.

\begin{figure}[]
\resizebox{0.5\columnwidth}{!}{\includegraphics{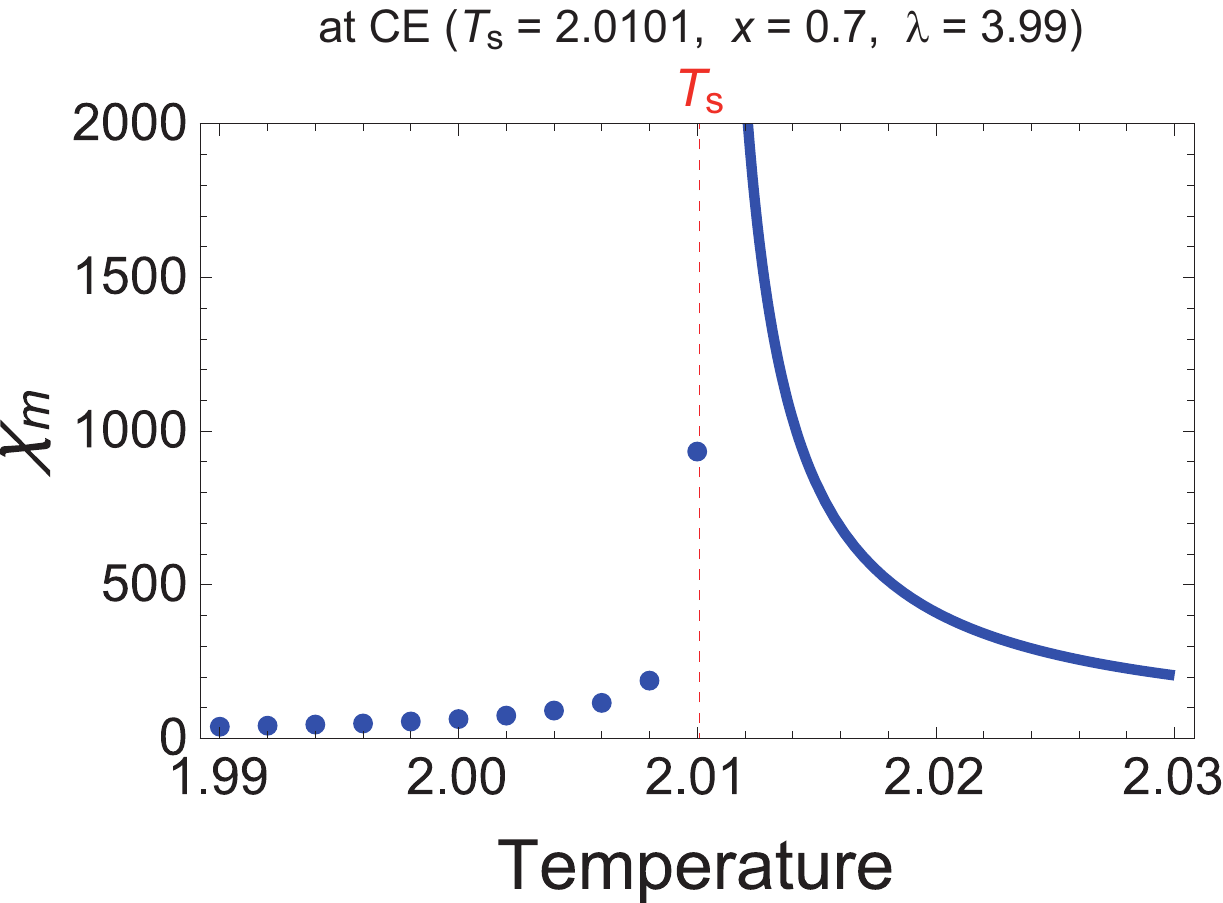}}
\caption{[color online] Susceptibility $\chi_m$ as a function of $T$. It diverges for $T>T_s$, whereas it becomes finite for $T<T_s$.} 
\label{fig:double_layer_network}
\end{figure}

\end{document}